\documentstyle[12pt]{article}

\def\beq{\begin{equation}} \def\eeq{\end{equation}}
\def\bea{\begin{eqnarray}} \def\eea{\end{eqnarray}}
\let\nn=\nonumber
\def\beann{\begin{eqnarray*}} \def\eeann{\end{eqnarray*}}

\newcommand{\append}[1]{\protect\stepcounter{section}
                        \setcounter{equation}{0} 
                        \section*{Appendix \thesection \qd #1}
                        \addcontentsline{toc}{section}{Appendix
                        \thesection: #1}}

\let\a=\alpha \let\be=\beta \let\g=\gamma \let\de=\delta
 \let\z=\zeta \let\h=\eta 
  \let\la=\lambda \let\m=\mu
\let\n=\nu  \let\p=\pi \let\r=\rho \let\s=\sigma
 
 \let\Ph=\phi  
 \let\Si=\Sigma 
 \let\G=\Gamma 

\let\qd=\quad  

\let\Ts=\textstyle  

\def\0{\over } \def\1{\vec }     \def\2{{1\over2}} \def\4{{1\over4}}
\def\5{\bar }  \def\6{\partial } \def\7#1{{#1}\llap{/}}

\def\<{\langle } \def\>{\rangle }  
 \let\LRA=\Leftrightarrow

\let\auf=\uparrow \let\ab=\downarrow

\def\CL{{\cal L}} \def\CR{{\cal R}}
\def\CT{{\cal T}}
\def\i{{\rm i}} \def\tr{\mbox{tr}} 
\def\str{\mbox{str}}

\def\sh{\mbox{\,sh}} \def\ch{\mbox{\,ch}}

\def\sn{\mbox{\,sn}} \def\cn{\mbox{\,cn}} \def\dn{\mbox{\,dn}}
 \def\res{\mbox{\,res}}

\newfont{\fettohne}{cmssbx10 scaled 1000}
\newfont{\gfettohne}{cmssbx10 scaled 1200}
\newfont{\kfettohne}{cmssbx10 scaled 700}

\def\MC{\mbox{{\gfettohne C}}}
\def\MZ{\mbox{{\gfettohne Z}}}

\def\sotimes{\hspace{2pt}
\mbox{\raisebox{2pt}{$\otimes$}}\hspace{-6.5pt}
\mbox{\raisebox{-5pt}{\sf s}}\hspace{4pt}}

\newcommand{\stensor}[2]{\left(#1\hspace{2pt}
\mbox{\raisebox{2pt}{$\otimes$}}\hspace{-6.5pt}
\mbox{\raisebox{-5pt}{\sf s}}\hspace{4pt}#2\right)}

\begin{document}

\thispagestyle{empty}
\begin{center}
{\Large {\bf Algebraic and Analytic Properties of the
One-Dimensional Hubbard Model\\}}
\vspace{7mm}
{\large Frank G\"{o}hmann$^\dagger$\footnote
{e-mail: frank@monet.phys.s.u-tokyo.ac.jp}
and Shuichi Murakami$^\ddagger$\footnote[4]
{e-mail: murakami@appi.t.u-tokyo.ac.jp}}\\
\vspace{5mm}
$^\dagger$Department of Physics, Faculty of Science,\\ University
of Tokyo,\footnote[5]{Address from Oct.\ 1996: Physikalisches
Institut der Universit\"at Bayreuth, TP1, 95440 Bayreuth,
Germany}\\
Hongo 7-3-1, Bunkyo-ku, Tokyo 113, Japan\\
$^\ddagger$Department of Applied Physics, Faculty of Engineering,\\
University of Tokyo,\\
Hongo 7-3-1, Bunkyo-ku, Tokyo 113, Japan
\vspace{7mm}

{\large {\bf Abstract}}
\end{center}
\begin{list}{}{\addtolength{\rightmargin}{10mm}
               \addtolength{\topsep}{-5mm}}
\item
We reconsider the quantum inverse scattering approach to the
one-dimensio\-nal Hubbard model and work out some of its basic
features so far omitted in the literature. It is our aim to show
that $R$-matrix and monodromy matrix of the Hubbard model, which
are known since ten years now, have good elementary properties.
We provide a meromorphic parametrization of the transfer matrix
in terms of elliptic functions. We identify the momentum
operator for lattice fermions in the expansion of the
transfer matrix with respect to the spectral parameter and
thereby show the locality and translational invariance of all
higher conserved quantities. We work out the transformation
properties of the monodromy matrix under the su(2) Lie algebra
of rotations and under the $\h$-pairing su(2) Lie algebra. Our
results imply su(2)$\oplus$su(2) invariance of the transfer
matrix for the model on a chain with an even number of sites.
\end{list}

\section{Introduction}
The one-dimensional Hubbard model is one of the most thoroughly
studied integrable quantum systems with applications in solid
state physics. Starting with the seminal article \cite{LiWu68}
of Lieb and Wu lots of its physical properties have been worked
out exactly \cite{EsKoBo}. For the case of half-filled band, in
particular, a complete picture of its elementary excitations is
available by now \cite{EsKo94a,EsKo94b}. All excited states are
scattering states of only four quasiparticles, two of which
carry spin but no charge, whereas the other two carry charge but
no spin. The $S$-matrix of these quasiparticles has been
calculated exactly. These achievements give a precise meaning to
the notion of spin-charge separation in one-dimensional solids.

All exact results on physical properties of the one-dimensional
Hubbard model obtained so far rely on the extensive use of the
coordinate Bethe Ansatz. Since Bethe wave functions, however,
are difficult to handle, any exact calculation of local
quantities going beyond the long distance asymptotics of
correlation functions \cite{FrKo91} seems to demand for an
algebraic treatment. Moreover, an algebraic treatment is likely
to facilitate the calculation of the thermodynamical properties
of the model \cite{KlBa96}.

At present we know two algebraic structures related to the
Hubbard model, a graded Yang-Baxter algebra, developed in the
works of Shastry \cite{Shastry86a,Shastry86b,Shastry88b} and
Olmedilla et al. \cite{WOA87,OWA87,OW88} and a representation
of the Y(su(2)) Yangian quantum group commuting with the Hubbard
Hamiltonian, which was discovered by Uglov and Korepin
\cite{UgKo94}. The relation of these two notions was recently
exposed by the authors \cite{MuGo96a}.

Although $R$-matrix and $L$-matrix of the Hubbard model are
known since long, it took nearly ten years before it was shown
that the $R$-matrix satisfies the Yang-Baxter equation
\cite{ShWa95}. An algebraic Bethe Ansatz was performed only
recently in a remarkable preprint by Ramos and Martins
\cite{RaMa96b}. Every progress in the development of an
algebraic approach was hindered before by the complexity of
$R$-matrix and monodromy matrix and by several unusual features
of these basic tools of the quantum inverse scattering method.
The monodromy matrix is $4 \times 4$ rather than $3 \times 3$,
as one might have guessed naively from the fact that there are
two levels of Bethe Ansatz equations. It further seems to be
impossible to find a parametrization of the $R$-matrix, such that
it becomes a function of the difference of the spectral
parameters.

Amazingly, not even the most elementary properties of $R$-matrix
and monodromy matrix have been worked out so far. For that
reason we look again at the construction of the graded
Yang-Baxter algebra. We begin with a description of the spin
model \cite{Shastry86a,Shastry86b, Shastry88b} that is related
to the Hubbard model by means of a Jordan-Wigner transformation
in section 2. Our account is based on the Yang-Baxter equation.
In section 3 we show that there exists a meromorphic parametrization
of the transfer matrix in terms of elliptic functions. Section 4
is devoted to the rederivation of the graded Yang-Baxter algebra
\cite{OWA87}. We show how to treat general twisted boundary
conditions. As a byproduct we find a simple method to obtain
higher conserved quantities of the Hubbard model from their
counterparts for the spin model. These conserved quantities are
generated by the graded trace of the fermionic monodromy matrix.
We identify the momentum operator for fermions in the zeroth
order of the expansion of this generating function with respect
to the spectral parameter. Thus all higher conserved quantities
are local and translational invariant. In section 5 we derive
the properties of the monodromy matrix under a combined
particle-hole and gauge transformation, which is
characteristical for the Hubbard Hamiltonian. It turns out that
the graded trace of the monodromy matrix is invariant under this
transformation, if the model is considered on a chain with an
even number of sites. In section 6 we investigate the behaviour of
the monodromy matrix under su(2) transformations. Our results are
complementary to the work of Ramos and Martins \cite{RaMa96b}
and should provide a means to discuss the symmetry properties
of quasiparticles within the algebraic approach. Like the
Hamiltonian, the graded trace of the monodromy matrix turns out
to be invariant under rotations of the spins and, if we consider
an even number of sites, also under the $\h$-pairing su(2) Lie
algebra \cite{HeLi71,Yang89,YaZh90,Pernici90}.
In order to make our representation self-contained, we expose the
$R$-matrix along with a list of relations among its elements in
appendix A. Appendix B provides a detailed discussion of the
momentum operator on a lattice. In appendix C it is shown how to
obtain the momentum operator from a monodromy matrix for free
fermions.
\section{The spin model}
As all one-dimensional fermionic models the Hubbard model is
related to a certain spin chain by a Jordan-Wigner
transformation. The Yang-Baxter algebra corresponding to this
spin chain is easier to formulate and closer to intuition
than the graded Yang-Baxter algebra of the Hubbard model. In
fact, Shastry in his seminal articles
\cite{Shastry86a,Shastry86b,Shastry88b} was using the language of
spins rather than the language of electrons. The graded form of
the Yang-Baxter algebra having the advantage of being defined
directly in terms of fermi operators was derived later by
Olmedilla et al.\ \cite{WOA87,OWA87,OW88}. Let us follow the
historical route here and start with a description of the spin
model. Its Hamiltonian is
\beq \label{hspin}
     H = \sum_{j=1}^L \left( \s_{j \tau}^+ \s_{j+1 \tau}^-
            + \s_{j \tau}^- \s_{j+1 \tau}^+ + {\Ts \frac{U}{4}}
              \, \s_{j \auf}^z \s_{j \ab}^z \right) \qd.
\eeq
Here and in the following we are using implicit summation over
doubly occuring indices. $H$ describes a periodic spin chain
of $L$ sites ($\s_{L+1 \tau}^{\pm} := \s_{1 \tau}^{\pm}$) with
two species of spins, labeled $\auf$ and $\ab$, at each site.
$U$ is the strength of an on-site Ising coupling between the
species. The interaction between nearest neighbours is of
XX-type for both species independently. Thus, in the limit
$U \rightarrow 0$ the model decouples into a pair of
non-interacting XX-chains.

As was shown by Shastry, $H$ is the logarithmic derivative of
a certain transfer matrix, which can be obtained by
appropriately coupling together two copies of the Yang-Baxter
algebra of the XX-chain. The $R$-matrix of the XX-chain is
\beq \label{rxx}
     r = {\Ts \frac{1}{2}} \left(a + b + (a - b)
               \s^z \otimes \s^z + \s^x \otimes \s^x +
               \s^y \otimes \s^y \right) \qd,
\eeq
where $a$ and $b$ have to satisfy the free fermion condition
\beq \label{freeab}
     a^2 + b^2 = 1 \qd.
\eeq
If we introduce the parametrization $a = \cos(\la)$, $b =
\sin(\la)$, then $r = r(\la)$ satisfies the Yang-Baxter equation
in the form
\beq \label{ybed}
     r_{12} (\la - \m) r_{13} (\la) r_{23} (\m) =
     r_{23} (\m) r_{13} (\la) r_{12} (\la - \m)
\eeq
which is sometimes called difference form of the Yang-Baxter
equation. $r(\la)$ is regular, i.e.\ $r(0) = P$, the permutation of
the two factors of $\MC^2 \otimes \MC^2$. As usual, the indices in
(\ref{ybed}) refer to the canonical embeddings of $r(\la)$ into
$\MC^2 \otimes \MC^2 \otimes \MC^2$. The $L$-matrix $l(\la)$ of
the XX-chain is the fundamental representation $l(\la) = r(\la)$
of the Yang-Baxter algebra,
\beq \label{xxlyba}
     \check r(\la - \m) (l(\la) \otimes l(\m)) =
         (l(\m) \otimes l(\la)) \check r(\la - \m) \qd,
\eeq
generated by $\check r(\la) := P r(\la)$. As usual,
$l(\la)$ in eq.\ (\ref{xxlyba}) has to be understood as matrix
in auxiliary space with entries acting on a quantum space.

For the construction of $R$- and $L$-matrices corresponding to
the spin Hamiltonian (\ref{hspin}) we have to duplicate the above
construction by attaching a label referring to the spin species
to each $\s$-matrix. The matrix $r(\la)$ is replaced by
$r_{\tau} (\la)$ ($\tau = \auf, \ab$), and we may redefine
$r(\la)$ as
\beq
     r(\la) := r_{\auf} (\la) r_{\ab} (\la) \qd.
\eeq
This new $R$-matrix obviously satisfies (\ref{ybed}) and is
again regular in the sense that $r(0) = P_{\auf} P_{\ab} =: P$
is a permutation operator.  Whenever an explicit matrix
representation is required we will use the convention
$\s_{\auf}^{\a} = \s^{\a} \otimes I_2$,
$\s_{\ab}^{\a} = I_2 \otimes \s^{\a}$, where $I_2$ denotes the
$2 \times 2$ unit matrix. Correspondingly, the $4 \times 4$
unit matrix, which will be needed below, is denoted by $I_4$.
Now Shastry's $R$-matrix associated to the Hamiltonian
(\ref{hspin}) reads
\beq \label{rspin}
     R(\la,\m|h,l) := \ch(h - l) \frac{r(\la - \m)}{\cos(\la - \m)}
           +  \sh(h - l) \frac{r(\la + \m)}{\cos(\la + \m)}
           \s^z \otimes \s^z \otimes I_4 \qd.
\eeq
This is a four parameter family of $16 \times 16$-matrices. It
was shown by Shiroishi and Wadati \cite{ShWa95}, that it
satisfies the Yang-Baxter equation in the form
\beq \label{yben}
     R_{12} (\la,\m|h,l) R_{13} (\la,\n|h,m) R_{23} (\m,\n|l,m)
        = R_{23} (\m,\n|l,m) R_{13} (\la,\n|h,m)
          R_{12} (\la,\m|h,l) \qd,
\eeq
provided that the parameters are constrained by the equations
\beq \label{const}
     \frac{U}{4} = \frac{\sh(2h)}{\sin(2\la)} =
                   \frac{\sh(2l)}{\sin(2\m)} =
                   \frac{\sh(2m)}{\sin(2\n)} \qd.
\eeq
This constraint can be satisfied for arbitrarily small $m$ and
$\n$. Hence,
\beq \label{lspin}
     L_{jk} (\la|h) := \cos(\la) R_{jk} (\la,0|h,0)
                     = r_{jk} (\la) e^{h \s_{j \auf}^z \s_{j \ab}^z}
\eeq
is a representation of the Yang-Baxter algebra generated by the
$R$-matrix (\ref{rspin}). The index $j$ in (\ref{lspin}) refers
to the auxiliary space, the index $k$ to the quantum space. Note
that $L_{jk} (\la|h) \ne L_{kj} (\la|h)$. $L_{jk} (\la|h)$ is a
tensor product of two XX-chain $L$-matrices coupled in auxiliary
space. If we solve the first equation in
(\ref{const}) for $h$ and choose the branch of solution
properly, then $h(\la = 0) = 0$, and $L_{jk}$ according to
(\ref{lspin}) is again regular. For $U = 0$ the constraint
(\ref{const}) is satisfied by $h = 0$ for all $\la$. We get back
the free model as it should be by construction, and the
$R$-matrix (\ref{rspin}) becomes a solution of the Yang-Baxter
equation in difference form (\ref{ybed}). On the other extreme,
we can carry out the limit $U \rightarrow \infty$ after properly
rescaling $\la \rightarrow \la/U$. Then the constraint
(\ref{const}) becomes $\la = 2 \sh(2h)$. However, $\la$
disappears from the definitions of $L_{jk}$ and $R_{jk}$, since
$\cos(\la/U) \rightarrow 1$ and $\sin(\la/U) \rightarrow 0$ for
every $\la$. Thus $r_{jk} ((\la - \m)/U) \rightarrow P_{jk}$, and
\beq \label{limuinf}
     R_{jk} (\la,\m|h,l) \rightarrow L_{jk} (0|h - l) =
        P_{jk} e^{(h - l)\s_{j \auf}^z \s_{j \ab}^z} \qd.
\eeq
It is easily verified that the expression on the right hand
side of (\ref{limuinf}) satisfies the Yang-Baxter equation in
difference form (\ref{ybed}) and that the corresponding Hamiltonian
is the on-site part of (\ref{hspin}) with $U = 1$. Hence $\la$ is
the natural spectral parameter of the model at zero coupling,
whereas $h$ is the natural spectral parameter in the strong
coupling limit. The algebra (\ref{rspin}), (\ref{yben}),
(\ref{const}) interpolates between these limits. For the
remainder of this article we will suppress the arguments $h$
and $l$ of the $R$- and $L$-matrices, assuming that they are given
as functions of $\la$ and $\m$ by the constraint (\ref{const}).

To finish the description of the spin model let us introduce its
monodromy matrix,
\beq \label{spinmono}
     T_L (\la) := L_{aL} (\la) \dots L_{a1} (\la) \qd.
\eeq
The index $a$ in this definition refers to the auxiliary space.
The transfer matrix of the spin model, $t(\la)$, is the
trace of $T_L (\la)$ over the auxiliary space. It follows from
the regularity of $L_{jk}$ that $\tilde U := t(0)$ is the shift
operator for spins,
\beq \label{tspin}
     \tilde U \s_{j \tau}^{\a} =
        \s_{j+1 \tau}^{\a} \tilde U \qd, \qd
        j = 1, \dots, L \qd, \qd \tau = \auf, \ab \qd.
\eeq
A brief calculation yields the derivative of the $L$-matrix
at zero spectral parameter,
\beq \label{locham}
     \dot{L}_{jk} (0) P_{jk} = \s_{j \tau}^+ \s_{k \tau}^- +
                               \s_{j \tau}^- \s_{k \tau}^+ +
                               {\Ts \frac{U}{4}} \s_{k \auf}^z
                               \s_{k \ab}^z \qd.
\eeq
This equation implies that the Hamiltonian (\ref{hspin}) is obtained
from the transfer matrix $t(\la)$ as logarithmic derivative,
$H = d_{\la} \left. \ln(t(\la)) \right|_{\la = 0}$.
\section{A meromorphic parametrization of the transfer matrix}
The considerations in this section were motivated by two facts.
First, in case of the eight vertex model Baxter's meromorphic
parametrization of the $R$-matrix yields a solution of
difference form of the Yang-Baxter equation \cite{Babook}.
Second, the transfer matrix enters certain functional
equations, the solutions of which usually require strong
analytic properties.

Let us solve the constraint (\ref{const}) for $e^{2h}$,
\beq
     e^{2h} = \frac{U \sin(2\la)}{4} +
              \sqrt{1 + \frac{U^2 \sin^2 (2\la)}{16}} \qd.
\eeq
The only possibility to remove the square root on the right hand
side is by replacing $2\la$ by $\mbox{am}(u)$, the
amplitude function and setting $k = \i U/4$. Then $e^{2h}$ is
expressed in terms of Jacobi elliptic functions as
\beq
     e^{2h} = \dn(u) - \i k \, \sn(u) \qd.
\eeq
Because of the homogeneity of the Yang-Baxter algebra, we may
multiply $L_{jk} (u)$ by $e^h$. Then $h(u)$ enters into the
definition of the monodromy matrix $T_a (u)$ only as
meromorphic function $e^{2h}$ of the redefined spectral
parameter $u$.  Letting $A := (a + b)/2$ and $B = (a - b)/2$
we see that the matrix $r_{jk}$ in the definition of the
$L$-matrix is of the form
\beq \label{oe}
     r_{jk} = \left( \begin{array}{cccc}
              e & o & o & e \\ o & e & e & o \\
              o & e & e & o \\ e & o & o & e
              \end{array} \right) \qd,
\eeq
where $e$ denotes an even polynomial in $A$, $B$ and $o$ denotes
an odd one. $e$ and $o$ are operators on quantum space $k$,
whose precise form is irrelevant for the following arguments.
The rules $e^2 = o^2 = e$, $ eo = oe = o$,
$e + e = e$ and $o + o = o$ imply
\beq
     \left( \begin{array}{cccc}
     e & o & o & e \\ o & e & e & o \\
     o & e & e & o \\ e & o & o & e
     \end{array} \right)^2 = 
     \left( \begin{array}{cccc}
     e & o & o & e \\ o & e & e & o \\
     o & e & e & o \\ e & o & o & e
     \end{array} \right) \qd.
\eeq
This means that the monodromy matrix (\ref{spinmono}) is again of
the form (\ref{oe}). Therefore $t(u)$ is an even polynomial in
$A$ and $B$. In other words, $t(u)$ is a  polynomial in $A^2$,
$B^2$ and $AB$. Since
\beq
     A^2 = (1 + \sn(u))/4 \qd, \qd
     B^2 = (1 - \sn(u))/4 \qd, \qd
      AB = \cn(u)/4 \qd,
\eeq
$t(u)$ is a meromorphic function of $u$.

To state the problem of the analytic structure of the $R$-matrix
let us go one step back and write again $a$ for $\cos(\la)$ and
$b$ for $\sin(\la)$. Let furthermore $c := e^{2h}$. Then $a$,
$b$ and $c$ are connected by the free fermion condition
(\ref{freeab}) and the constraint (\ref{const}), i.e.\ $a$, $b$
and $c$ are lying on a complex curve given by the algebraic
equations
\beq
     a^2 + b^2 = 1 \qd, \qd c - 1/c = U ab \qd.
\eeq
This curve may be called the spectral curve of the Hubbard
model. Unfortunately we could neither find a meromorphic
parametrization of this seemingly simple structure nor assign a
geometrical meaning to it.
\section{The Hubbard model}
Applying a Jordan-Wigner transformation to $R$- and $L$-matrix of
the spin model we obtain the graded Yang-Baxter algebra
\cite{KuSk82} of the Hubbard model \cite{OWA87}. No extra effort
is necessary to introduce the grading. It is rather induced by the
Jordan-Wigner transformation. Before turning to the fermionic
formulation let us perform a gauge transformation of $R$- and
$L$-matrices with $4 \times 4$ transformation matrix
\beq
     G_x := e^{x (\s^z \otimes \s^z)/2} \qd.
\eeq
Then
\bea \label{lschlange}
     L_k (\la) \qd \longrightarrow \qd \tilde{L}_k (\la) & := &
         G_h L_k (\la) G_h^{-1} = G_h
         (l_{k \auf} (\la) \otimes l_{k \ab} (\la)) G_h \qd, \\[1ex]
     \label{rschlange}
     R(\la, \m) \qd \longrightarrow \qd \tilde{R} (\la, \m) & := &
         (G_h \otimes G_l) R(\la, \m) (G_h^{-1} \otimes G_l^{-1})
         \qd.
\eea
We suppress the auxiliary space index of the $L$-matrix here and
in the following and consider $\tilde{L}_k (\la)$ as $4 \times 4$
matrix with entries acting on quantum space $k$. Unlike
$R(\la, \m)$, the transformed matrix $\tilde{R} (\la, \m)$ is
symmetric. By use of the definition
\beq \label{rcheck}
     \check{R} (\la, \m) := P \tilde{R} (\la, \m)
\eeq
the Yang-Baxter algebra assumes the form
\beq \label{symyba}
     \check{R} (\la, \m)
        \left( \tilde{L}_k (\la) \otimes \tilde{L}_k (\m) \right)
        = \left( \tilde{L}_k (\m) \otimes \tilde{L}_k (\la) \right)
          \check{R} (\la, \m) \qd,
\eeq
which is most convenient for changing to a fermionic formulation
\cite{KuSk82}.

Since the canonical anticommutation relations for fermi operators
are invariant under gauge transformations and under particle-hole
transformations, there is some freedom in the definition of the
Jordan-Wigner transformation. Using the abbreviations
\beq
     u_k := \sum_{j=1}^k n_{j \auf} \qd, \qd
     d_k := \sum_{j=1}^k n_{j \ab} \qd, k = 1, \dots , L,
\eeq
where the $n_{j s}$ are electron densities and setting further
$u_0 := d_0 := 1$, we take the choice
\bea \label{jwd1}
     \s_{k \auf}^+ = c_{k \auf}^+ e^{\i \p u_{k-1}} \qd &,& \qd
     \s_{k \auf}^- = c_{k \auf} e^{- \i \p u_{k-1}} \qd, \\[1ex]
     \label{jwd2}
     \s_{k \ab}^+ = c_{k \ab}^+ e^{\i \p (u_L + d_{k-1})} \qd &,& \qd
     \s_{k \ab}^- = c_{k \ab} e^{- \i \p (u_L + d_{k-1})} \qd.
\eea
The XX $L$-matrices $l_{k \auf} (\la)$ and $l_{k \ab} (\la)$
can now be expressed in terms of fermi operators,
\bea \label{lauf}
     l_{k \auf} (\la) & = & e^{- \i \p u_k \s^z /2}
                            \CL_{k \auf} (\la)
                            e^{\i \p u_{k-1} \s^z /2} \qd, \\[1ex]
     \label{lab}
     l_{k \ab} (\la) & = & e^{- \i \p (u_L + d_k) \s^z /2}
                           \CL_{k \ab} (\la)
                           e^{\i \p (u_L + d_{k-1}) \s^z /2} \qd,
\eea
where $\CL_{k \tau} (\la)$ ($\tau = \auf, \ab$) is defined as
\beq \label{freef}
     \CL_{k \tau} (\la) := \left( \begin{array}{cc}
                       \sin(\la) + \i e^{\i \la} n_{k\tau} &
                       c_{k\tau} \\ - \i c_{k\tau}^+ &
                       \cos(\la) - e^{\i \la} n_{k\tau}
                       \end{array} \right) \qd.
\eeq
$\CL_{k \tau} (\la)$ is an $L$-matrix for free fermions. Its
entries for different quantum space indices $k$ either commute or
anticommute. This fact can be formally described by assigning a
parity $\p(\CL_{k \tau} (\la)_j^i) = 0, 1$ to each matrix
element. Call a matrix element even, if its parity is zero, odd
if it is one. Odd matrix elements with different quantum space
indices anticommute, whereas even elements commute with all
elements with different quantum space indices. A grading is a
function $p$, which assigns parity to the basis vectors of
the auxiliary space. Let $p(i) := p(e_i)$. Then, with the
grading $p(1) = 0$, $p(2) = 1$, $p \in Z_2$, the elements of
the free fermion $L$-matrix have parity
$\p (\CL_{k \tau} (\la)_j^i) = p(i) + p(j)$. The crucial point
about these notions is that they allow for the introduction of a
graded tensor product $\sotimes$ which respects comultiplication.
Let $\sotimes$ be defined by the equation
\beq \label{stensor}
     \stensor{A}{B}_{kl}^{ij} := (-1)^{(p(i) + p(k))p(j)}
                                 A_k^i B_l^j \qd.
\eeq
Then $\stensor{A}{B} \stensor{C}{D} = \stensor{AC}{BD}$ for all
matrices $A, B, C, D$ of the parity defined above. The graded
tensor product can be used to formulate a graded Yang-Baxter
algebra \cite{KuSk82}. Grading and graded tensor product may be
defined for arbitrary matrix dimensions.

In appendix C we give a brief account of the free fermion model
which will be needed below to recover the lattice momentum
operator from the monodromy matrix of the Hubbard model. The
free fermion monodromy matrix is
\beq \label{freemono}
     \CT_{L \tau} (\la) :=
        \CL_{L \tau} (\la) \dots \CL_{1 \tau} (\la) \qd, \qd
        (\tau = \auf, \ab) \qd.
\eeq
Note that our choice of Jordan-Wigner transformation and thus
our free fermion $L$-matrix differ from that in \cite{OWA87}.

Inserting (\ref{lauf}), (\ref{lab}) into (\ref{lschlange}) we
find
\beq \label{slhl}
     \tilde{L}_k (\la) = W V_k^{-1} \CL_k (\la) V_{k-1} W^{-1}
     \qd,
\eeq
where
\bea
     V_k & := & e^{\i \p u_k \s^z /2} \otimes
                e^{\i \p (u_L + d_k) \s^z /2} \qd, \\[1ex]
     \label{W}
     W & := & \mbox{diag} (1,1,\i,\i) \qd, \\[1ex]
     \label{lhub}
     \CL_k (\la) & = &
        G_h \stensor{\CL_{k \auf}(\la)}{\CL_{k \ab}(\la)} G_h \qd.
\eea
The grading comes in naturally in (\ref{lhub}), since the
operator $I_2 \otimes e^{- \i\pi (u_L + d_k) \s^z /2}$ when moved
to the left in the tensor product $l_{k \auf} (\la) \otimes
l_{k \ab} (\la)$ induces a gauge transformation on $\CL_{k \auf}
(\la)$, which affects only the odd elements. We will see below
that $\CL_k (\la)$ according to equation (\ref{lhub}) is an
$L$-matrix for the Hubbard model. The associated $R$-matrix
follows from (\ref{symyba}). First of all we have
\bea \label{lltrans}
     \lefteqn{\tilde{L}_k (\la) \otimes \tilde{L}_k (\m) =}
     \nn \\[1ex] &&
       (W \otimes W) \left( V_k^{-1} \otimes V_k^{-1} \right)
       X \stensor{\CL_k (\la)}{\CL_k (\m)} X^{-1}
       \left( V_{k-1} \otimes V_{k-1} \right)
       (W^{-1} \otimes W^{-1}) \,, \qd
\eea
where $\sotimes$ is a graded tensor product of $4 \times 4$
matrices (cf.\ (\ref{stensor})) with grading $p(1) = p(4) = 0$,
$p(2) = p(3) = 1$. We are using the same symbol for
graded tensor products of $2 \times 2$ and $4 \times 4$
matrices. For $4 \times 4$ matrices the grading will always be
as above, and for $2 \times 2$ matrices we will use the grading
introduced below (\ref{stensor}). The matrix $X$ in
(\ref{lltrans}) is the diagonal matrix
\beq
     X := \s^z \otimes \mbox{diag} (1,\i,\i,1) \otimes I_2 \qd.
\eeq
Let
\beq \label{gamma}
     \G := \mbox{diag} (e^{\i \a}, e^{\i \be},
                           e^{\i \g}, e^{\i \de}) \qd,
\eeq
where $\a, \be, \g, \de$ may generally be mutually commuting
operators. Then
\beq \label{rcom}
     \left[ \check R (\la, \m), \G \otimes \G \right] = 0 \qd,
\eeq
\beq \label{abcd}
     \LRA \qd \a + \de = \be + \g \, \mbox{mod} 2 \p \qd.
\eeq
Since $V_k$ and W are of the form (\ref{gamma}) and satisfy
(\ref{abcd}), we may infer from (\ref{lltrans}) that
\beq \label{hubyba}
     \CR (\la, \m) \stensor{\CL_k (\la)}{\CL_k (\m)} =
        \stensor{\CL_k (\m)}{\CL_k (\la)} \CR (\la, \m) \qd,
\eeq
where
\beq \label{hubr}
     \CR (\la, \m) = X^{-1} \check R (\la, \m) X \qd.
\eeq
Hence, the $L$-matrix of the Hubbard model is a representation
of the graded Yang-Baxter algebra. This result is due to
Olmedilla et al.\ \cite{OWA87}. To be self-contained we expose
the $R$-matrix $\CR (\la,\m)$ along with some useful relations
among its elements in appendix A.

The graded tensor product in (\ref{hubyba}) respects
comultiplication. Therefore the monodromy matrix
\beq
     \CT_L (\la) := \CL_L (\la) \dots \CL_1 (\la)
\eeq
represents the graded Yang-Baxter algebra with the same $R$-matrix
\beq \label{hubtyba}
     \CR (\la, \m) \stensor{\CT_L (\la)}{\CT_L (\m)} =
        \stensor{\CT_L (\m)}{\CT_L (\la)} \CR (\la, \m) \qd.
\eeq
We will demonstrate below that $\CT_L (\la)$ generates the
Hubbard Hamiltonian. For the matrix elements of $\CT_L (\la)$
we introduce the following notation
\beq \label{mblock}
     \CT_L (\la) = \left( \begin{array}{cccc}
                   D_{11} & C_{11} & C_{12} & D_{12} \\
                   B_{11} & A_{11} & A_{12} & B_{12} \\
                   B_{21} & A_{21} & A_{22} & B_{22} \\
                   D_{21} & C_{21} & C_{22} & D_{22}
                   \end{array} \right) \qd,
\eeq
dividing it into four $2 \times 2$ submatrices $A(\la)$, $B(\la)$,
$C(\la)$, $D(\la)$. As we shall see in sections 5 and 6, this block
notation reflects the properties of the monodromy matrix under
the two su(2) transformations connected with the Hubbard model
and under combined particle-hole and gauge transformations.

If $\a$, $\be$, $\g$, $\de$ satisfy (\ref{abcd}) and commute
among each other and with the diagonal elements of $\CT_L (\la)$,
then (\ref{rcom}), (\ref{hubr}) and (\ref{hubtyba}) imply that
\beq
     [\tr(\G \, \CT_L (\la)), \tr(\G \, \CT_L (\m))] = 0 \qd.
\eeq
Thus $\tr (\G \, \CT_L (\la))$ generates a family of mutually
commuting operators. Different choices of $\a$, $\be$, $\g$,
$\de$ correspond to different boundary conditions. Since we did
not restrict $\a$, $\be$, $\g$, $\de$ to be complex numbers, a
dynamical twist is possible. In fact, because of the non-local
nature of the Jordan-Wigner transformation, the periodic spin
model turns into the Hubbard model with dynamically twisted
boundary conditions. This point was recently discussed in detail
by Yue and Deguchi \cite{YuDe96}. To express the transfer matrix
$t(\la)$ of the spin model introduced in section 2 in terms of
fermi operators we note that $\G = V_L^{-1}$ satisfies
(\ref{abcd}). Using (\ref{lschlange}) and (\ref{slhl}) we
conclude that
\beq
     t(\la) = \tr (V_L^{-1} \CT_L (\la)) \qd.
\eeq
As we will see in the following, the Hubbard model under
periodic boundary conditions is obtained with the choice
$\G = \s^z \otimes \s^z$, which leads to
\beq \label{hubt}
     \str (\CT_L (\la)) := \tr ((\s^z \otimes \s^z) \CT_L (\la))
        = \tr(D) - \tr(A) = \str (V_L T_L (\la)) \qd.
\eeq
This expression is called the graded trace or super trace of the
monodromy matrix. Its zeroth order term of the expansion in the
spectral parameter is
\beq \label{hubshift}
     \str (\CT_L (0)) = \str (\CT_{L \auf} (0))
                        \str (\CT_{L \ab} (0))
        = e^{- \i \p u_L /2} \hat{U}_{\auf} e^{- \i \p d_L /2}
          \hat{U}_{\ab}
        = e^{- \i \p \hat{N} /2} \hat U \qd.
\eeq
where $\hat N = u_L + d_L$ is the particle number operator, and
$\hat U$ is the shift operator for electrons. (\ref{hubshift})
follows from the corresponding result for free fermions which
we derive in appendix C. $\hat U$ is the product
$\hat U = \hat U_{\auf} \hat U_{\ab}$ of shift operators for up
and down spin electrons. We introduce these operators in appendix
B, where we give a detailed account of shift and momentum
operators for fermions on the lattice. $\hat U$ is connected
to the total momentum $\Pi$ by $ \hat U = e^{\i \Pi}$. $\Pi$
assumes its familiar form,
\beq \label{pi}
     \Pi = \Ph \sum_{k=1}^{L-1} k \tilde{c}_{k \tau}^+
                                  \tilde{c}_{k \tau} \qd,
\eeq
when expressed in terms of Fourier transformed fermi operators,
\beq
     \tilde{c}_{k \tau} = \frac{1}{\sqrt{L}} \sum_{l=1}^L
                          e^{\i \Ph kl} c_{l \tau} \qd.
\eeq
For brevity we wrote $\Ph := 2\p/L$ here. Eq.\ (\ref{pi}) has to
be read with care, since $\Pi$ is merely defined modulo $2 \pi$.
Hence we interpret (\ref{pi}) as defining equation of an
equivalence class of operators differing from each other only by
certain ``phase operators''. A restriction $\hat \Pi$ of $\Pi$ to
the fundmental domain of the logarithm is constructed in appendix
B. $\hat \Pi$ is a polynomial in $\hat U$.

The momentum operator $\hat \Pi$ preserves the particle number.
Thus
\beq
     \ln(\str(\CT_L (0))) = - \i \p \hat N /2 + \i \hat \Pi \qd.
\eeq
We will see moreover in section 6 that $\str(\CT_L (\la))$
commutes with the particle number operator and may therefore
conclude that
\beq \label{pcom}
     [\hat \Pi, \ln(\str(\CT_L (\la)))] = 0 \qd.
\eeq
This equation implies that
$\tau (\la) := \ln(\str(\CT_L (\la)))$ is a generating
function of translational invariant commuting operators.
According to the arguments of L\"uscher \cite{Luescher76}
these operators are local. They are most easily calculated in
the language of the spin model, since the building blocks of the
monodromy matrix are permutation operators of spins,
$P_{jk \tau} = \Ts{\2} (1 + \s_{j \tau}^\a \s_{k \tau}^\a)$.

To give an example, we present the derivation of the
Hamiltonian. Recall from section 2 that $\tilde U = t(0)$
is the shift operator for spins. If we reintroduce
for a moment an auxiliary space index $a$, we obtain
\beq
     \dot T_{aL} (0) = P_{a1} \tilde U \dot L_{L L-1} (0) P_{L L-1}
                       + \sum_{j=1}^{L-1}
                       \dot L_{j+1 j} (0) P_{j+1 j} P_{a1} \tilde U
                       \qd.
\eeq
The product $\dot L_{jk} (0) P_{jk}$ was given in section 2,
eq.\ (\ref{locham}). Using the Jordan-Wigner transformation,
(\ref{jwd1}), (\ref{jwd2}), it is expressed in terms of fermi
operators as
\beq
     \dot L_{j+1 j} (0) P_{j+1 j} =
        c_{j \tau}^+ c_{j+1 \tau} + c_{j+1 \tau}^+ c_{j \tau} +
        U(n_{j \auf} - \Ts{\frac{1}{2}})
         (n_{j \ab} - \Ts{\frac{1}{2}}) \qd.
\eeq
(\ref{hubt}) and (\ref{hubshift}) imply
$\str(V_{a L} P_{a1} \tilde U) = e^{- \i \p \hat N /2} \hat U$.
Since moreover $[V_{aL},\dot L_{j+1 j} (0) P_{j+1 j}] = 0$, we
obtain $\str(\dot \CT_{aL} (0)) = e^{- \i \p \hat N /2}
\hat U \hat H$. 

$\hat H$ is the Hubbard Hamiltonian
\beq \label{hhub}
     \hat H = \sum_{j=1}^L \left(
        c_{j \tau}^+ c_{j+1 \tau} + c_{j+1 \tau}^+ c_{j \tau} +
        U(n_{j \auf} - \Ts{\frac{1}{2}})
         (n_{j \ab} - \Ts{\frac{1}{2}}) \right)
\eeq
under periodic boundary
conditions ($c_{L+1 \tau} := c_{1 \tau}$). Due to our choice of
Jordan-Wigner transformation (\ref{jwd1}), (\ref{jwd2}) we
obtained the Hamiltonian for holes here. The sign of the hopping
term can of course be changed by a particle-hole transformation.
Higher conserved quantities may be calculated in the same way as
the Hamiltonian. We get an expansion of the generating function
$\tau (\la)$, whose first terms are
\beq \label{tauexp}
     \tau(\la) = - \i \p \hat N /2 + \i \hat \Pi + \la \hat H +
                   {\cal O} (\la^2) \qd.
\eeq
The ${\cal O} (\la^2)$ term was derived by Shastry
\cite{Shastry88b}. The zeroth order terms in (\ref{tauexp}) were
not known before. They are however indispensable for the
derivation of the dispersion relations of elementary excitations
from the eigenvalues of $\str(\CT_L (\la))$
\cite{RaMa96b,YuDe96}. It will be interesting to investigate, if
we can get both branches of quasiparticle dispersion relations
\cite{EsKo94a,EsKo94b} from these eigenvalues.
\section{Discrete transformations}
A characteristic feature of the Hubbard Hamiltonian on a chain
consisting of an even number of sites is its invariance under
the transformation
\beq \label{disctrans}
     c_{j \auf} \rightarrow c_{j \auf} \qd, \qd
     c_{j \ab} \rightarrow (-1)^j c_{j \ab}^+ \qd, \qd
     U \rightarrow - U \qd.
\eeq
Since the generators of rotations of the spins are not invariant
under (\ref{disctrans}), there exists a second su(2) Lie algebra
commuting with the Hamiltonian.

We will show now that not only the Hamiltonian, but the whole
transfer matrix $\str(\CT_L (\la))$ is invariant under
(\ref{disctrans}). First note that $h(\la) \rightarrow - h(\la)$,
and thus
\beq
     G_h \rightarrow G_{-h} = G_h^{-1} \qd.
\eeq 
The matrix elements of the free fermion $L$-matrix (\ref{freef})
transform according to
\beq
     \CL_{k \ab} (\la) \rightarrow
        e^{\i \p k \s^z /2} \s^y \CL_{k \ab} (\la)
        \s^y e^{- \i \p (k -1) \s^z /2} \qd.
\eeq
The last two formulae imply
\beq
     \CL_k (\la) \rightarrow
        \left( I_2 \otimes e^{\i \p k \s^z /2} \right)
        (\s^z \otimes \s^y) \CL_k (\la) (\s^z \otimes \s^y)
        \left( I_2 \otimes e^{- \i \p (k - 1) \s^z /2} \right)
        \qd,
\eeq
and thus by comultiplication,
\beq \label{transtauf}
     \CT_L (\la) \rightarrow
        \left( I_2 \otimes e^{\i \p L \s^z /2} \right)
        (\s^z \otimes \s^y) \CT_L (\la) (\s^z \otimes \s^y) \qd.
\eeq
Finally $\str(\CT_L (\la))$ transforms according to
\beq
     \str(\CT_L (\la)) \rightarrow - e^{- \i \p L/2} \left(
        D_{11} (\la) - e^{\i \p L} A_{11} (\la) - A_{22} (\la)
        + e^{\i \p L} D_{22} (\la) \right) \qd,
\eeq
and we can conclude invariance modulo sign of the graded trace
for even $L$
\beq \label{transinv}
     \str(\CT_L (\la)) \rightarrow \pm \str(\CT_L (\la)) \qd.
\eeq
Hence, all higher commuting operators generated by $\tau (\la) =
\ln(\str(\CT_L (\la)))$ are invariant under the transformation
(\ref{disctrans}).

We can of course revers the spins in
(\ref{disctrans}). Then a slight modification in the
transformation of the monodromy matrix occurs. The factors $\s^z
\otimes \s^y$ in (\ref{transtauf}) have to be replaced by $\s^y
\otimes I_2$, and the two factors of $I_2 \otimes e^{\i \p L \s^z /2}$
have to be interchanged. The result (\ref{transinv})
for the transfer matrix remains the same. Performing up spin and
down spin transformations in succession the monodromy matrix
transforms as
\beq
     \CT_L (\la) \rightarrow \left(
         e^{\i \p L \s^z /2} \otimes e^{\i \p L \s^z /2} \right)
         (\s^x \otimes \s^y) \CT_L (\la) (\s^x \otimes \s^y) \qd.
\eeq
In this case we find for the transfer matrix
\beq
     \str(\CT_L (\la)) \rightarrow
        \tr(e^{- \i \p L \s^z} D (\la)) - \tr(A(\la)) \qd.
\eeq
Again invariance is only achieved for an even number of lattice
sites.
\section{su(2) symmetries}
A careful discussion of the two su(2) symmetries
\cite{HeLi71,Yang89,YaZh90,Pernici90} connected with the Hubbard
Hamiltonian was crucial for the complete understanding of the
coordinate Bethe Ansatz of the model. It was shown in
\cite{EKS92a} that the states obtained from coordinate Bethe
Ansatz are incomplete. They are highest weight states of two
su(2) Lie algebras \cite{EKS91,EKS92b}. One is the su(2) Lie
algebra of rotations, the other one is the $\h$-pairing su(2) Lie
algebra. The generators of the $\h$-pairing su(2) are obtained
from the generators of rotations under the canonical
transformation of the preceding section. They are connected with
the creation of pairs of particles or holes in the system.

We show in the following how the generators of the two
symmetries commute with the monodromy matrix. Our result will be
useful for the classification of quasiparticles according to
their symmetry within the algebraic approach. A discussion
analogous to the discussion of the spin of spin waves by Faddeev
and Takhtajan \cite{TaFa84} is likely to be possible. There are
four interactionless states which may serve as reference states
for an algebraic Bethe Ansatz of the Hubbard model, the empty
band, the completely filled band and the half-filled band with
all spins up or all spins down. Depending on the choice of
reference state four of the elements of the matrices $B(\la)$
and $C(\la)$ in (\ref{mblock}) are creation operators, whereas
the remaining four are annihilation operators. This fits with the
fact that there are four different quasiparticles in the
system. We think that their identification will eventually become
possible by use of the symmetry properties presented below.

The su(2) generators of rotations are given by
\beq \label{glogen1}
     S^+ :=  - \sum_{j=1}^L c_{j \auf}^+ c_{j \ab} \qd, \qd
     S^- :=  - \sum_{j=1}^L c_{j \ab}^+ c_{j \auf} \qd, \qd
     S^z :=  \sum_{j=1}^L (n_{j \auf} - n_{j \ab}) \qd.
\eeq
Recall that we are using the language of holes here (cf.\
(\ref{hhub})). Under a particle hole transformation the
operators $S^+$, $S^-$, $S^z$ turn into the operators
$\z^\dagger$, $\z$, $\z_z$ used by E{\ss}ler et al.\
\cite{EsKo94b}.  We will show now that the whole transfer
matrix is rotational invariant. To this end let us introduce
local generators of rotations
\beq
     S_j^+ := - c_{j \auf}^+ c_{j \ab} \qd, \qd
     S_j^- := - c_{j \ab}^+ c_{j \auf} \qd, \qd
     S_j^z := n_{j \auf} - n_{j \ab} \qd.
\eeq
The matrices
\beq
     \Si^+ := \s^+ \otimes \s^- \qd, \qd
     \Si^- := \s^- \otimes \s^+ \qd, \qd
     \Si^z := \Ts{\frac{1}{2}} (\s^z \otimes I_2 - I_2 \otimes \s^z)
\eeq
clearly generate a representation of su(2). They are connected to
the inner block $A(\la)$ of the monodromy matrix $\CT_L (\la)$.
Let
\bea
     \Si^x & := & \Si^+ + \Si^- \qd, \qd
     \Si^y := - \i (\Si^+ - \Si^-) \qd, \\[1ex]
     S_j^x & := & S_j^+ + S_j^- \qd, \qd
     S_j^y := - \i (S_j^+ - S_j^-) \qd.
\eea
Then it is not difficult to see that
\beq \label{locsu2}
     [\CL_j (\la),\Si^\a + S_j^\a] = 0 \qd, \qd \a = x, y, z \qd.
\eeq
The verification of this equation may be done as follows. First
show by direct calculation that $[\CL_j (\la),\Si^+ + S_j^+] = 0$.
$\CL_j (\la)$ has the same block structure as the
monodromy matrix, (\ref{mblock}). Under reversing all spins the
blocks $A(\la)$, $B(\la)$, $C(\la)$, $D(\la)$ transform as
\beq
     \left( \begin{array}{cc} A & B \\ C & D \end{array} \right)
     \rightarrow
     \left( \begin{array}{cc} \s^x & 0 \\ 0 & \s^z \end{array}
     \right)
     \left( \begin{array}{cc} A & B \\ C & D \end{array} \right)
     \left( \begin{array}{cc} \s^x & 0 \\ 0 & \s^z \end{array}
     \right) \qd,
\eeq
and we may conclude that $[\CL_j (\la),\Si^- + S_j^-] = 0$. The
vanishing of the last commutator $[\CL_j (\la), \Si^z + S_j^z]$
follows by means of the Jacobi identity.

The local equation (\ref{locsu2}) extends to an identity for the
monodromy matrix by induction,
\beq \label{rotmono}
     [\CT_L (\la),\Si^\a + S^\a] = 0 \qd,
\eeq
$\a = x, y, z$, where $S^x $ and $S^y$ are defined as their
local analogs. Taking the graded trace of this equation yields
\beq
     [\str(\CT_L (\la)), S^\a] = 0 \qd, \qd \a = x, y, z \qd.
\eeq
The transfer matrix and thus all higher commuting operators are
rotational invariant.

The transformation properties of the monodromy matrix under the
discrete transformation (\ref{disctrans}) introduced in the
preceding section induce a second su(2) invariance. Applying
(\ref{disctrans}) to the su(2) generators of rotations,
(\ref{glogen1}), we find
\bea
     S^+ & \rightarrow & \h^+ = \sum_{j=1}^L
        (-1)^{j+1} c_{j \auf}^+ c_{j \ab}^+ \qd, \\[1ex]
     S^- & \rightarrow & \h^- = \sum_{j=1}^L
        (-1)^{j+1} c_{j \ab} c_{j \auf} \qd, \\[1ex]
     \label{etaz}
     S^z & \rightarrow & \h^z = \sum_{j=1}^L
        (n_{j \auf} + n_{j \ab} - 1) = \hat N - L \qd.
\eea
This is the $\h$-pairing symmetry. Because of
(\ref{etaz}), it may be interpreted as non-abelian extension of
the gauge symmetry. The commutators of the generators of
$\h$-pairing with the monodromy matrix follow from
(\ref{disctrans}) and (\ref{rotmono}). Let
\beq
     \tilde \Si^+ := \s^+ \otimes \s^+ \qd, \qd
     \tilde \Si^- := \s^- \otimes \s^- \qd, \qd
     \tilde \Si^z := \Ts{\frac{1}{2}}
                      (\s^z \otimes I_2 + I_2 \otimes \s^z) \qd.
\eeq
These matrices generate a representation of su(2) connected to
the block $D(\la)$ of the monodromy matrix. Like in case of
rotations we define
\bea
     \tilde \Si^x & := & \tilde \Si^+ + \tilde \Si^- \qd, \qd
     \tilde \Si^y := - \i (\tilde \Si^+ - \tilde \Si^-) \qd, \\[1ex]
     \h^x & := & \h^+ + \h^- \qd, \qd
     \h^y := - \i (\h^+ - \h^-) \qd.
\eea
Using these definitions we obtain for $L$ even
\beq \label{etamono}
     [\CT_L (\la), \tilde \Si^\a + \h^\a] = 0
        \qd, \qd \a = x, y, z \qd,
\eeq
and thus
\beq \label{etastr}
     [\str(\CT_L (\la)),\h^\a] = 0 \qd.
\eeq
Note that there is no local analog like (\ref{locsu2}) of
equation (\ref{etastr}). The $\h$-pairing symmetry is sensitive
to a change of boundary conditions and is in this sense a non-local
symmetry.

Equation (\ref{etamono}) may be verified in the following way.
Observe that
\beq \label{si1}
     \Si^\pm (\s^z \otimes \s^y) =
        (\s^z \otimes \s^y) \tilde \Si^\pm \qd,
\eeq
and
\beq \label{si2}
     \Si^\pm (I_2 \otimes e^{\i \p L \s^z /2}) =
        (I_2 \otimes e^{- \i \p L \s^z /2}) \Si^\pm \qd.
\eeq
Apply the transformation (\ref{disctrans}) to equation
(\ref{rotmono}), and use (\ref{si1}), (\ref{si2}). Then
\beq \label{etapmmono}
     [\CT_L (\la),\h^\pm] + \CT_L (\la) \tilde \Si^\pm -
        (I_2 \otimes e^{\i \p L \s^z})
        \tilde \Si^\pm \CT_L (\la) = 0 \qd.
\eeq
Equation (\ref{si2}) remains true, if $\Si^\pm$ is replaced by
$\tilde \Si^\pm$. Using this fact, (\ref{etapmmono}) implies
\beq \label{etazmono}
     [\CT_L (\la),\tilde \Si^z + \h^z] = 0 \qd,
\eeq
whereby
\beq
     [\str(\CT_L (\la)),\hat N] = 0
\eeq
for every $L$, which means that all higher conserved quantities
are gauge invariant. This fact has been used in the derivation
of (\ref{pcom}). From (\ref{etapmmono}) and (\ref{etazmono}) we
infer the validity of (\ref{etamono}) for even $L$.

Here is a simple example for the usefulness of the above
formulae. (\ref{tauexp}), (\ref{etaz}), (\ref{etastr}) imply
immediately that $\hat \Pi \h^+ = \h^+ (\hat \Pi + \p)$, which
means that $\h^+$ changes the momentum of eigenstates by $\p$
\cite{Yang89}.
\section{Conclusions}
We hope we could convince the reader, that the graded
Yang-Baxter algebra (\ref{hubtyba}) is a useful tool for further
investigations of the one-dimensional Hubbard model. Our account
of basic features of the monodromy matrix should be read in
conjunction with the recent preprint of Ramos and Martins
\cite{RaMa96b}, who were able to diagonalize $\str(\CT_L (\la))$
by purely algebraic means. Combining both works it should be not
too difficult to redrive algebraically all results obtained so
far by means of the coordinate Bethe Ansatz. Moreover, there is
hope to proceed in the calculation of correlation functions and
thermodynamical properties.

There have been speculations \cite{EsKo94a} that there might be a
different Yang-Baxter algebra embedding of the Hubbard
Hamiltonian. Of course we can not rule out this possibility.
Some arguments in favour of it, however, are disproved by now.
In a recent article \cite{MuGo96a} we were able to show how a
rational substructure of the $R$-matrix naturally arises in the
thermodynamic limit. The corresponding submatrix of the monodromy
matrix generates the Y(su(2)) representation discovered by Uglov
and Korepin \cite{UgKo94}. This nicely fits with the fact that
the $S$-matrix of quasiparticle scattering is of rational form.
In the present article we showed that the monodromy matrix has
an appropriate algebraic structure. In particular, its graded
trace is fully su(2)$\oplus$su(2) invariant and invariant under
translations.  The analytic properties of $R$-matrix and monodromy
matrix are less usual. We do not yet have a geometrical idea of the
spectral curve. Still we succeeded in showing the existence of a
meromorphic parametrization of the transfer matrix.

{\bf Acknowledgments}. This  work has been supported by the Japan
Society for the Promotion of Science and the Ministry of Science,
Culture and Education of Japan. We are grateful to Professor
Miki Wadati for continuous encouragement and comments. We would
like to thank H.\ Fehske, V.\ E.\ Korepin and M.\ Shiroishi for
fruitful discussions. S.\ M.\ is also grateful to Professor
Naoto Nagaosa for his encouragement.

\renewcommand{\thesection}{\Alph{section}}
\renewcommand{\theequation}{\thesection.\arabic{equation}}
\setcounter{section}{0}
\clearpage

\append{The R-matrix}
The $R$-matrix generating the graded Yang-Baxter algebra of
the Hubbard model was first derived by Olmedilla et al.\
\cite{OWA87}. It follows from the equations (\ref{rspin}),
(\ref{rschlange}), (\ref{rcheck}) and (\ref{hubr}) and is of
the following structure,
\setlength{\arraycolsep}{2pt}
\bea
     \lefteqn{\CR (\la,\m) =} \nn \\[1ex]
        && \left( \begin{array}{cccccccccccccccc}
           \r_1 &0&0&0 &0&0&0&0 &0&0&0&0 &0&0&0&0 \\
           0& \r_2 &0&0 & \i \r_9 &0&0&0 &0&0&0&0 &0&0&0&0 \\
           0&0& \r_2 &0 &0&0&0&0 & \i \r_9 &0&0&0 &0&0&0&0 \\
           0&0&0& \r_3  &0&0& - \i \r_6 &0 &0& \i \r_6 &0&0
           & \r_8 &0&0&0 \\
           0& -\i \r_{10} &0&0 & \r_2 &0&0&0 &0&0&0&0 &0&0&0&0 \\
           0&0&0&0 &0& \r_4 &0&0 &0&0&0&0 &0&0&0&0 \\
           0&0&0& \i \r_6  &0&0& \r_5 &0 &0& \r_7 &0&0
           & - \i \r_6 &0&0&0 \\
           0&0&0&0 &0&0&0& \r_2 &0&0&0&0 &0& - \i \r_{10} &0&0 \\
           0&0& - \i \r_{10} &0 &0&0&0&0 & \r_2 &0&0&0 &0&0&0&0 \\
           0&0&0& - \i \r_6  &0&0& \r_7 &0 &0& \r_5 &0&0
           & \i \r_6 &0&0&0 \\
           0&0&0&0 &0&0&0&0 &0&0& \r_4 &0 &0&0&0&0 \\
           0&0&0&0 &0&0&0&0 &0&0&0& \r_2 &0&0& - \i \r_{10} &0 \\
           0&0&0& \r_8  &0&0& \i \r_6 &0 &0& - \i \r_6 &0&0
           & \r_3 &0&0&0 \\
           0&0&0&0 &0&0&0& \i \r_9 &0&0&0&0 &0& \r_2 &0&0 \\
           0&0&0&0 &0&0&0&0 &0&0&0& \i \r_9 &0&0& \r_2 &0 \\
           0&0&0&0 &0&0&0&0 &0&0&0&0 &0&0&0& \r_1
        \end{array} \right) \\ \nn
\eea
The ten Boltzmann weights $\r_j = \r_j (\la, \m)$ are
\bea
     \r_1 & = & \cos(\la) \cos(\m) e^{h - l} +
            \sin(\la) \sin(\m) e^{l - h} \qd, \\[1ex]
     \r_2 & = & 1 \qd, \\[1ex]
     \r_3 & = & \frac{\cos(\la) \cos(\m) e^{h - l} -
            \sin(\la) \sin(\m) e^{l - h}}
            {\cos^2 (\la) - \sin^2 (\m)} \qd, \\[1ex]
     \r_4 & = & \cos(\la) \cos(\m) e^{l - h} +
            \sin(\la) \sin(\m) e^{h - l} \qd, \\[1ex]
     \r_5 & = & \frac{\cos(\la) \cos(\m) e^{l - h} -
            \sin(\la) \sin(\m) e^{h - l}}
            {\cos^2 (\la) - \sin^2 (\m)} \qd, \\[1ex]
     \r_6 & = & \frac{2 \sh(2(h - l))}
            {U (\cos^2 (\la) - \sin^2 (\m))} \qd, \\[1ex]
     \r_7 & = & \r_4 - \r_5 \qd, \\[1ex]
     \r_8 & = & \r_1 - \r_3 \qd, \\[1ex]
     \r_9 & = & \sin(\la) \cos(\m) e^{l - h} -
            \cos(\la) \sin(\m) e^{h - l} \qd, \\[1ex]
     \r_{10} & = & \sin(\la) \cos(\m) e^{h - l} -
            \cos(\la) \sin(\m) e^{l - h} \qd.
\eea
The parameters $\la$, $\m$, $h$ and $l$ are connected by equation
(\ref{const}). Note that our definition of $h$ and $l$ differs
from that in ref.\ \cite{OWA87} and that we performed a shift of
$\frac{\p}{4}$ in the arguments of the functions $\a (\la)$  and
$\g (\la)$ occurring there.

There are the following quadratic relations between the
Boltzmann weights, \cite{OWA87},
\bea
     \r_1 \r_4 + \r_9 \r_{10} & = & 1 \qd, \\[1ex]
     \r_1 \r_5 + \r_3 \r_4 & = & 2 \qd, \\[1ex]
     \r_3 \r_5 - \r_6^2 & = & 1 \qd.
\eea
Further identities useful for practical calculations can be
found in the recent article \cite{MuGo96a}.
\append{Momentum operator for fermions on a lattice}
Below we present a detailed discussion of the
momentum operator on a lattice.
\subsection*{The shift operator from permutations}
In this subsection we mimic the construction of the shift operator
for spin chains. Start with spinless fermions, $c_1, \dots, c_L$ on
a one dimensional lattice of $L$ sites. Let
\beq \label{deftrans}
     K_{ij} := 1 - (c_i^+ - c_j^+)(c_i - c_j) \qd.
\eeq
It is not difficult to see that $K_{ij}$ permutes fermions. There
are the obvious identities
\beq \label{e1}
     K_{ij} = K_{ij}^+ \qd, \qd K_{ij} = K_{ji} \qd, \qd
     K_{jj} = 1 \qd.
\eeq
Use of the fundamental anti commutators for the fermions
yields
\beq \label{perc}
     K_{ij} c_i = c_i + (c_i^+ - c_j^+) c_j (c_i - c_j)
                = c_i + (- 1 + c_j c_j^+ - c_j c_i^+)
                            (c_i - c_j) = c_j K_{ij} \qd,
\eeq
and (\ref{e1}) implies
\beq
     K_{ij} c_j = c_i K_{ij} \qd, \qd
     K_{ij} c_i^+ = c_j^+ K_{ij} \qd, \qd
     K_{ij} c_j^+ = c_i^+ K_{ij} \qd.
\eeq
Furthermore, by the last four equations, we obtain
\beq \label{trans}
    K_{ij} K_{jk} = K_{ik} K_{ij} = K_{jk} K_{ik} \qd, \qd
                    i \ne j \ne k \qd,
\eeq
and a short calculation similar to the one in eq.\ (\ref{perc})
leads to
\beq \label{qua}
    K_{ij} K_{ij} = 1 \qd.
\eeq
Hence, the operators $K_{ij}$ are identified as permutation
operators.

With the aid of permutations of fermions it is of course
possible to realize a global shift \cite{HeLi71}. Let
\beq \label{defU}
     \hat U := K_{12} K_{23} \dots K_{L-1 L} \qd.
\eeq
Then eq.\ (\ref{perc}) implies
\beq \label{shift}
     \hat U c_j = \left\{
     \begin{array}{l}
         c_{j+1} \hat U \\[1ex]
         c_{1} \hat U
     \end{array}
     \qd, \qd \mbox{if} \qd
     \begin{array}{l}
         j = 1, \dots, L -1 \\[1ex]
         j = L \qd.
     \end{array}
     \right.
\eeq
This means that $\hat U$ is acting as right shift operator on
the elementary fermi operators of a periodic chain of $L$ sites.
Now (\ref{defU}) implies
\beq
     \hat U^+ = K_{L-1 L} \dots K_{12} \qd,
\eeq
and thus by eq.\ (\ref{qua}), $\hat U \hat U^+ =
\hat U^+ \hat U = 1$.  $\hat U$ is invertible, and
$\hat U^{-1} = \hat U^+$, i.e.\ $\hat U$ is unitary.
$\hat U^{-1}$ is the left shift operator. With the aid of the
shift operator it is possible to define the momentum as its
formal infinitesimal generator,
\beq \label{defP}
     e^{\i \Pi} := \hat U \qd.
\eeq
This is the common construction in case of the spin chains.
Note however, that (\ref{defP}) defines $\Pi$ only modulo
$2 \pi$.

To realize the shift operator for electrons, attach a spin label
to all operators above, and observe that
$[\hat U_\auf, c_{j \ab}] = [\hat U_\ab, c_{j \auf}] = 0$. Then
$\hat U := \hat U_\auf \hat U_\ab$ is the shift operator for
electrons.
\subsection*{Diagonalization of the momentum operator -- an
exercise}
For simplicity, return to the spinless case. Because of the
basic property (\ref{shift}), it follows that $[\hat U^L,c_j] =
[\hat U^L,c_j^+] = 0$, $j = 1, \dots, L$, i.e.\ $\hat U^L$ is a scalar
operator. Since $\hat U|0\> = |0\>$, we obtain $\hat U^L = 1$. Shifting
all fermions once around the lattice does not change the state
of the system. This simple fact has strong implications.

Let $\a \in \MC$. Then
\beq
     (1 - \a \hat U) \, \sum_{n=0}^{L-1} \a^{n} \hat U^{n} =
        1 - \a^{L} \qd.
\eeq
This means that the resolvent of $\hat U$ is a finite sum. With
$\la := 1/\a$ we find,
\beq \label{reso}
     (\la - \hat U)^{-1} = \frac{1}{\la (\la^L - 1)}
                     \sum_{n=0}^{L-1} \la^{L-n} \hat U^n \qd.
\eeq
The sum on the right hand side of the latter equation is regular
in $\la$ and of order $\la$. The spectrum of $\hat U$ is therefore
given by the equation
\beq
     \la^L = 1 \qd, \qd \LRA \qd, \qd \la_{k} = e^{\i 2\p k/L}
               \qd, \qd k = 0,\dots, L - 1 \qd,
\eeq
and is of course highly degenerate.

Let $P_0,\dots, P_{L-1}$ be the projections on the eigenspaces
of $\hat U$ corresponding to the eigenvalues $\la_0,\dots, \la_{L-1}$.
The $P_k$ are orthogonal, since the eigenstates of a unitary
operator to different eigenvalues are. $P_k P_l = \de_{kl} P_k$,
and furthermore, $\sum_{k=0}^{L-1} P_k = 1$. Hence, the spectral
decomposition of the resolvent (\ref{reso}) is
\beq
     (\la - \hat U)^{-1} = \sum_{k=0}^{L-1} \frac{P_k}{\la - \la_k}
                      \qd.
\eeq
Conversely, the $P_k$ are determined by the spectral
decomposition via $P_{k} = \res_{\la = \la_k}
((\la - \hat U)^{-1})$. Since
$d_{\la} (\la (\la^L - 1))|_{\la = \la_k} = L$, we obtain
\beq
     P_k = \frac{1}{L} \sum_{n=0}^{L-1} e^{- \i \Ph kn} \hat U^n \qd,
\eeq
where we used the abbreviation $\Ph = 2\p/L$. From this
representation the momentum eigenstates arise quite naturally.
The particle number operator $\hat{N}$ is of course translational
invariant, $[\hat{N},\hat U] = 0$. Hence, the projections
$P_k$ preserve the particle number, $[P_k,\hat{N}] = 0$. We find
\beq
     P_k c_j^+ |0\> = \frac{e^{\i \Ph jk}}{L} \sum_{n=1}^L
                     e^{- \i \Ph kn} c_n^+ |0\> \qd.
\eeq
Since this is true for all $j = 1,\dots, L$, we see that $\hat U$ is
nondegenerate in the one particle sector of the Hilbert space.
In this sector the subspace corresponding to the eigenvalue
$\la_k$ is spanned by the right hand side of the above equation.
Normalization yields the eigenvector $|k\> := \tilde{c}_k^+ |0\>$,
where $\tilde{c}_k^+$ is the creator of a one particle momentum
eigenstate. It is defined in the usual way as the adjoint of the
annihilator
\beq \label{defct}
     \tilde{c}_k := \frac{1}{\sqrt{L}}
                    \sum_{n=1}^{L} e^{\i \Ph kn} c_n \qd.
\eeq
$\tilde{c}_k$ and $\tilde{c}_k^+$ are fermi operators. They are
transformed back into site operators $c_j$, $c_j^+$ by Fourier
inversion of eq.\ (\ref{defct}). Therefore arbitrary operators
defined in terms of $c_j$, $c_j^+$ may be expressed in terms of
$\tilde{c}_k$ and $\tilde{c}_k^+$. The particle number operator,
in particular, is form invariant under Fourier transformation,
\beq \label{Nt}
     \hat{N} = \sum_{k=0}^{L-1} \tilde{c}_k^+ \tilde{c}_k \qd.
\eeq
Since the $\tilde{c}_k$ are fermi operators, the states
$|k_1 \dots k_N\> := \tilde{c}_{k_1}^+ \dots \tilde{c}_{k_N}^+
|0\>$, $k_1 < \dots < k_N$, are orthogonal. (\ref{Nt})
implies $\hat{N}|k_1 \dots k_N\> = N |k_1 \dots k_N\>$. Counting
these states we see that they span the $N$-particle sector of the
lattice Hilbert space. Letting $N$ vary from $1,\dots, L$ we
get a basis of the full Hilbert space.

Now, from the definition (\ref{defct}) of $\tilde{c}_k$ it
follows that
\beq \label{smo}
     \hat U \tilde{c}_k^+ = e^{\i \Ph k} \tilde{c}_k^+ \hat U \qd.
\eeq
Hence we have obtained
\beq
     \hat U|k_1 \dots k_N\> = e^{\i \Ph (k_1 + \dots + k_N)}
                         |k_1 \dots k_N\> \qd,
\eeq
or for the momentum operator, respectively,
\beq
     \Pi |k_1 \dots k_N\> = \Ph (k_1 + \dots + k_N)
                           |k_1 \dots k_N\> \qd.
\eeq
This equation is of course defined only modulo $2\pi$.
On the other hand it is easily verified that the operator
$\Ph \sum_{k=1}^{L-1} k \, \tilde{c}_k^+ \tilde{c}_k$ acts the same
way on $|k_1 \dots k_N\>$. Since these states form a basis, we
have achieved the diagonalization of the momentum operator,
\beq \label{pidiag}
     \Pi = \Ph \sum_{k=1}^{L-1} k \, \tilde{c}_k^+ \tilde{c}_k \qd.
\eeq
In this form $\Pi$ is usually found in the literature.
\subsection*{Site representation of the momentum operator}
What happens, when we use the Fourier transform to translate back
the above diagonal form of the momentum operator into the site
representation? Inserting the definition (\ref{defct}) into
eq.\ (\ref{pidiag}) yields
\beq \label{pre}
     \Pi = \frac{\Ph}{L} \sum_{m,n = 1}^L c_m^+ c_n
           \sum_{k=1}^{L-1} k e^{- \i \Ph (m - n)k} \qd.
\eeq
Now, for $\a \in \MC$, let
\beq \label{defg}
     g(\a ) := \sum_{k=0}^{L-1} \i e^{- \i \Ph k \a} =
               \i \, \frac{1 - e^{- \i \Ph L \a}}
                          {1 - e^{- \i \Ph \a}} \qd,
\eeq
where the term on the right hand side makes sense only for $\a
\notin L \MZ$. Eqs.\ (\ref{pre}) and (\ref{defg}) imply that
\bea \nn
     \Pi & = & \frac{1}{L} \sum_{m,n = 1}^L g'(m - n) \, c_m^+ c_n
               \\[1ex] \label{palt}
         & = & {\Ts \2} \Ph (L - 1) \hat{N} +
               \Ph \sum_{m,n = 1 \atop m \ne n}^L
               \frac{c_m^+ c_n}{e^{- \i \Ph (m - n)} - 1} \qd.
\eea
This formula is remarkable. Since $\Pi = - \i \ln (\hat U)$, we can
read it as having taken the logarithm of the ordered product of
transpositions which defines $\hat U$. $\Pi$ as given by
(\ref{palt}) is a non-local operator. It is interesting to
notice that the second term on the right hand side of
(\ref{palt}) is gauge equivalent to the $1/\sin(\Ph (m - n))$
hopping term of the long-range Hubbard model introduced by
Gebhard and Ruckenstein \cite{GeRu92}. We have thus found a
simple interpretation of this Hamiltonian and in turns a simple
interpretation of the origin of the $1/\sin^2$ exchange of the
Haldane-Shastry spin chain \cite{Haldane88,Shastry88a}.
\subsection*{Momentum operator modulo $2 \pi$}
To obtain the appropriate definition of $\Pi$ mod $2\p$, observe
from the preceding subsection that
\beq \label{zahn}
     \Ph k = \frac{1}{L} \sum_{m=1}^L g'(m) e^{\i \Ph km}
           = \Ph \sum_{m=1}^{L-1} \left( \2 + 
             \frac{e^{\i \Ph km}}{e^{- \i \Ph m} -1} \right)
\eeq
for $k = 0,\dots,L - 1$. With view of the right hand side of this
equation the function $\Ph k$ is periodically continued to a saw
tooth function on the integers. Since $\Pi /\Ph$ assumes only
integer eigenvalues, the definition
\beq
     \hat \Pi := \Ph \sum_{m=1}^{L-1} \left( \2 +
                 \frac{\hat U^m}{e^{- \i \Ph m} - 1} \right)
\eeq
yields the required restriction of $\Pi$ modulo $2\pi$. In other
words, (\ref{smo}) and (\ref{zahn}) imply that $e^{\i \Pi} =
e^{\i \hat \Pi}$. $\hat \Pi$ obviously commutes with the Hubbard
Hamiltonian, whereas $\Pi$ does not.
\append{Free fermions}
In this appendix we show how to obtain the shift operator $\hat
U$ for fermions from the graded trace of the free fermion
monodromy matrix (\ref{freemono}). For this purpose we have to
treat XX-chain and free fermion model in parallel. Both models
are related by a Jordan-Wigner transformation. For brevity let
us consider the up-spin case (\ref{jwd1}), and let us suppress
the spin index here. The XX-chain monodromy matrix is
\beq \label{xxmono}
     T_L (\la) = l_L (\la) \dots l_1 (\la) \qd.
\eeq
It satisfies the Yang-Baxter algebra (\ref{xxlyba}),
\beq \label{xxmyba}
     \check r (\la - \m) (T_L (\la) \otimes T_L (\m)) =
        (T_L (\m) \otimes T_L (\la)) \check r (\la - \m) \qd.
\eeq
Applying the Jordan-Wigner transformation (\ref{jwd1}) to the
XX-chain monodromy matrix, we obtain
\beq \label{tt}
     T_L (\la) = e^{- \i \p u_L \s^z /2} \CT_L (\la) \qd,
\eeq
where $\CT_L (\la)$ is the free fermion monodromy matrix
(\ref{freemono}) with $\tau = \auf$, which satisfies the graded
analog of (\ref{xxmyba}),
\beq \label{freeyba}
     \check r_g (\la - \m) \stensor{\CT_L (\la)}{\CT_L (\m)} =
        \stensor{\CT_L (\m)}{\CT_L (\la)} \check r_g (\la - \m)
        \qd.
\eeq
The matrix $\check r_g (\la)$ is defined as
$\check r_g (\la) := W^{-1} \check r(\la) W$ with $W$ according
to (\ref{W}). (\ref{freeyba}) implies that
\beq
     [\str(\CT_L (\la)),\str(\CT_L(\m))] = 0 \qd.
\eeq

We want to calculate the action of $\str(\CT_L(0))$ on fermi
operators. To this end let
\beq
     P_{0j} := {\Ts \2} (1 + \s^\a \s_j^\a) \qd, \qd
     P_{jk} := {\Ts \2} (1 + \s_j^\a \s_k^\a) \qd.
\eeq
Then (\ref{xxmono}) implies that
\beq
     T_L (0) = P_{01} P_{12} \dots P_{L-1 L} = P_{01} \tilde U \qd.
\eeq
We use the Jordan-Wigner
transformation (\ref{jwd1}) to express $P_{j j+1}$ in terms of
fermi operators. Then
\beq
     P_{j j+1} = K_{j j+1} + 2 n_j n_{j+1}
\eeq
with the spinless densities $n_j = c_j^+ c_j$ and the
permutation operators $K_{j j+1}$ according to (\ref{deftrans}).
Let $e_\pm := e^{\pm \i \p u_L/2}$. Then (\ref{tt}) implies that
\beq \label{strtr}
     \str(\CT_L (0)) = (e_+ n_1 + e_- (n_1 - 1)) \tilde U \qd.
\eeq
Using the Jordan-Wigner transformation (\ref{jwd1}) and the
known action of the permutation operators $P_{j j+1}$ on Pauli
matrices, we obtain
\bea
     \tilde U c_k & = & c_{k+1} (1 - 2n_1) \tilde U \qd, \qd
        k = 1, \dots , L-1 \qd, \\[1ex]
     \tilde U c_L & = & c_1 (1 - 2n_1) e_+^2 \tilde U \qd.
\eea
$e_+$ and $e_-$ are generators of gauge transformations,
\beq \label{egt}
     e_+ c_k = - \i c_k e_+ \qd, \qd e_- c_k = \i c_k e_- \qd.
\eeq
By use of the four previous formulae, we infer for the
annihilators $c_k$,
\beq \label{strc}
     \str(\CT_L(0)) c_k = \i c_{k+1} \str(\CT_L(0)) \qd,
\eeq
where $k = 1, \dots, L$ ($c_{L+1} = c_1$). It is not difficult to
see that $\str(\CT_L(0))$ is a unitary operator. Therefore
the creators $c_k^+$ satisfy
\beq \label{strcp}
     \str(\CT_L(0)) c_k^+ = - \i c_{k+1}^+ \str(\CT_L(0)) \qd,
\eeq
$k = 1, \dots, L$. (\ref{egt}), (\ref{strc}) and (\ref{strcp})
imply that
\beq
     e_+ \str(\CT_L (0)) = \a_L \hat U \qd,
\eeq
where $\a_L$ is a complex constant, and $\hat U$ is the shift
operator (\ref{defU}) for spinless fermions.
Since $K_{j j+1} |0\> = P_{j j+1} |0\> = |0\>$, we conclude
from (\ref{strtr}) that $\a_L = - 1$ and eventually obtain
\beq \label{ende}
     \str(\CT_L(0)) = - e_- \hat U \qd.
\eeq
The corresponding formula for down-spins is simply obtained by
reversing the spins. Then (\ref{ende}) implies (\ref{hubshift}).
It may be instructive for the reader to verify (\ref{ende})
directly for small $L$.

%\bibliographystyle{phys}
%\bibliography{hub}

\begin{thebibliography}{10}

\bibitem{LiWu68}
E.~H. Lieb and F.~Y. Wu, Phys. Rev. Lett. {\bf 20}, 1445 (1968).

\bibitem{EsKoBo}
F.~H.~L. E{\ss}ler and V.~E. Korepin, {\em Exactly Solvable
Models of Strongly Correlated Electrons}, World Scientific,
Singapore,  (1994).

\bibitem{EsKo94a}
F.~H.~L. E{\ss}ler and V.~E. Korepin,
Phys. Rev. Lett. {\bf 72}, 908 (1994).

\bibitem{EsKo94b}
F.~H.~L. E{\ss}ler and V.~E. Korepin,
Nucl. Phys. B {\bf 426}, 505 (1994).

\bibitem{FrKo91}
H.~Frahm and V.~E. Korepin, Phys. Rev. B {\bf 43}, 5653 (1991).

\bibitem{KlBa96}
A.~Kl\"{u}mper and R.~Z. Bariev, Nucl. Phys. B {\bf 458}, 623 (1996).

\bibitem{Shastry86a}
B.~S. Shastry, Phys. Rev. Lett. {\bf 56}, 1529 (1986).

\bibitem{Shastry86b}
B.~S. Shastry, Phys. Rev. Lett. {\bf 56}, 2453 (1986).

\bibitem{Shastry88b}
B.~S. Shastry, J. Stat. Phys. {\bf 50}, 57 (1988).

\bibitem{WOA87}
M.~Wadati, E.~Olmedilla, and Y.~Akutsu,
J. Phys. Soc. Jpn. {\bf 56}, 1340 (1987).

\bibitem{OWA87}
E.~Olmedilla, M.~Wadati, and Y.~Akutsu,
J. Phys. Soc. Jpn. {\bf 56}, 2298 (1987).

\bibitem{OW88}
E.~Olmedilla and M.~Wadati, Phys. Rev. Lett. {\bf 60}, 1595 (1988).

\bibitem{UgKo94}
D.~B. Uglov and V.~E. Korepin, Phys. Lett. A {\bf 190}, 238 (1994).

\bibitem{MuGo96a}
S.~Murakami and F.~G\"ohmann.
\newblock Yangian Symmetry and Quantum Inverse Scattering Method
for the One-Dimensional Hubbard Model.
\newblock preprint, cond-mat/9609249,  (1996).

\bibitem{ShWa95}
M.~Shiroishi and M.~Wadati, J. Phys. Soc. Jpn. {\bf 64}, 57 (1995).

\bibitem{RaMa96b}
P.~B. Ramos and M.~J. Martins.
\newblock Algebraic Bethe Ansatz Approach for the One-Dimensional
Hubbard Model.
\newblock preprint, hep-th/9605141,  (1996).

\bibitem{HeLi71}
O.~J. Heilmann and E.~H. Lieb,
Ann.~N.Y.~Acad.~Sci. {\bf 172}, 584 (1971).

\bibitem{Yang89}
C.~N. Yang, Phys. Rev. Lett. {\bf 63}, 2144 (1989).

\bibitem{YaZh90}
C.~N. Yang and S.~C. Zhang, Mod. Phys. Lett. B {\bf 4}, 40 (1990).

\bibitem{Pernici90}
M.~Pernici, Europhys. Lett. {\bf 12}, 75 (1990).

\bibitem{Babook}
R.~J. Baxter, {\em Exactly Solved Models in Statistical
Mechanics}, Academic Press, London,  (1982).

\bibitem{KuSk82}
P.~P. Kulish and E.~K. Sklyanin, J. Soviet Math. {\bf 19}, 1596 (1982).

\bibitem{YuDe96}
R.~Yue and T.~Deguchi.
\newblock Analytic Bethe Ansatz for 1-d Hubbard model and
twisted coupled XY model.
\newblock preprint, hep-th/9605141,  (1996).

\bibitem{Luescher76}
M.~L\"{u}scher, Nucl. Phys. B {\bf 117}, 475 (1976).

\bibitem{EKS92a}
F.~H.~L. E{\ss}ler, V.~E. Korepin, and K.~Schoutens,
Nucl. Phys. B {\bf 372}, 559 (1992).

\bibitem{EKS91}
F.~H.~L. E{\ss}ler, V.~E. Korepin, and K.~Schoutens,
Phys. Rev. Lett. {\bf 67}, 3848 (1991).

\bibitem{EKS92b}
F.~H.~L. E{\ss}ler, V.~E. Korepin, and K.~Schoutens,
Nucl. Phys. B {\bf 384}, 431 (1992).

\bibitem{TaFa84}
L.~A. Takhtajan and L.~D. Faddeev, J. Soviet Math. {\bf 24}, 241 (1984).

\bibitem{GeRu92}
F.~Gebhard and A.~E. Ruckenstein, Phys. Rev. Lett. {\bf 68}, 244 (1992).

\bibitem{Haldane88}
F.~D.~M. Haldane, Phys. Rev. Lett. {\bf 60}, 635 (1988).

\bibitem{Shastry88a}
B.~S. Shastry, Phys. Rev. Lett. {\bf 60}, 639 (1988).

\end{thebibliography}

\end{document}